\documentstyle[elsart12,osa,epsf,eqsecnum,epsfig,graphics,cite]{revtex}

\def\lsim{\raise0.3ex\hbox{$<$\kern-0.75em\raise-1.1ex\hbox{$\sim$}}}
\def\gsim{\raise0.3ex\hbox{$>$\kern-0.75em\raise-1.1ex\hbox{$\sim$}}}

\begin{document}


\voffset1.5cm


\title{Low-$p_T$ Collective Flow Induces High-$p_T$ Jet Quenching}
\author{N\'estor Armesto, Carlos A. Salgado and Urs Achim Wiedemann}

\address{Department of Physics, CERN, Theory Division,
CH-1211 Gen\`eve 23, Switzerland}
\date{\today}
\maketitle

\begin{abstract}
Data on low-$p_T$ hadronic spectra are widely regarded as evidence 
of a hydrodynamic expansion in nucleus-nucleus collisions. In this 
interpretation, different hadron species emerge from a common medium 
that has built up a strong collective velocity field. Here, we show 
that the existence of a collective flow field implies characteristic
modifications of high-$p_T$ parton fragmentation. 
We generalize the formalism of parton energy loss
to the case of flow-induced, oriented momentum transfer. We also
discuss how to embed this calculation in hydrodynamic
simulations. Flow effects are found to result generically
in characteristic asymmetries in the $\eta \times \phi$-plane of 
jet energy distributions and of multiplicity distributions associated 
to high-$p_T$ trigger particles. But collective 
flow also contributes to the medium-induced suppression of single inclusive 
high-$p_T$ hadron spectra. In particular, we find that 
low-$p_T$ elliptic flow can induce a sizeable additional  
contribution to the high-$p_T$ azimuthal asymmetry by selective elimination
of those hard partons which propagate with significant inclination
against the flow field. 
This reduces at least partially the recently observed problem
that models of parton energy loss tend to underpredict the 
large azimuthal asymmetry $v_2$ of high-$p_T$ hadronic spectra
in semi-peripheral Au+Au collisions.  
\end{abstract}

\section{Introduction}
\label{sec1}

What happens if a hard process, such as the production 
of high-$E_T$ jets, is embedded in a dense nuclear environment
created e.g. in a nucleus-nucleus collision at RHIC or at the
LHC ? While parton-parton interactions at high virtuality
$Q^2 \gg \Lambda_{QCD}^2$ occur on too short time and
length scales to be affected by the typical modes in the medium,
the parton showers associated to the incoming and
outgoing state can interact with the 
medium~\cite{Gyulassy:1993hr,Baier:1996sk,Zakharov:1997uu,Wiedemann:2000za,Gyulassy:2000er,Wang:2001if}.
This is expected to result in an energy degradation
of the leading parton~\cite{Gyulassy:1993hr,Baier:1996sk,Zakharov:1997uu,Wiedemann:2000za,Gyulassy:2000er,Wang:2001if}, in a transverse momentum broadening
of the parton shower~\cite{Wiedemann:2000tf,Baier:2001qw,Salgado:2003gb,Salgado:2003rv}, and in an enhanced and softened
multiplicity distribution of the hadronic final state~\cite{Salgado:2003rv}.
Measurements in Au+Au collisions at RHIC support this 
picture by the observed suppression of leading hadron 
spectra~\cite{Adcox:2001jp,Adler:2003au,Adler:2002xw,Adams:2003kv,Back:2003qr,Arsene:2003yk}
and leading back-to-back correlations~\cite{Adler:2002tq}, as well as the
medium-modified ``jet-like'' properties of particle
production associated to high-$p_T$ trigger 
particles~\cite{Wang:2004kf,Adler:2004zd,magestro}. 
The analysis of these ``jet quenching'' observables
has become one of the most active and most diverse research
fields in ultrarelativistic nucleus-nucleus collisions, 
mainly because the pattern of medium-induced partonic energy
loss is expected to allow for a detailed characterization of
the properties of the produced dense medium.

Parton energy loss is known to be
sensitive to the total in-medium pathlength and to the average
squared transverse momentum transferred from the medium to the
hard parton~\cite{Baier:2000mf,Kovner:2003zj,Gyulassy:2003mc,Jacobs:2004qv,Salgado:2003qc}. 
In recent phenomenological studies, the latter 
quantity has been parameterized by the BDMPS transport coefficient 
$\hat{q}$~\cite{Salgado:2003rv,Eskola:2004cr} or by physically
equivalent model-dependent quantities such as twist-4 multiple 
scattering matrix elements~\cite{Wang:2003aw}, the medium opacity or 
the number of initially produced gluons per unit 
rapidity~\cite{Gyulassy:2000gk}. These model parameters can be
related to the energy density of the produced 
matter~\cite{Baier:2002tc,Accardi:2003gp,Salgado:2003rv}:
\begin{equation}
  \hat{q} (\xi) = c\, \epsilon^{3/4}(\xi)\, .
  \label{1.1}
\end{equation}
Here, $c$ is a medium-dependent constant ($c \sim 2$ for the
case of an ideal quark gluon plasma~\cite{Baier:2002tc,Eskola:2004cr}), 
and (\ref{1.1}) also holds for a time-dependent energy density as 
indicated by the explicit $\xi$-dependence. 
Recently, several phenomenological analysis have used the medium 
modifications of high-$p_T$ hadron production to extract information about 
the energy density attained in nucleus-nucleus collisions at 
RHIC~\cite{Wang:2003aw,Eskola:2004cr,Dainese:2004te,Gyulassy:2003mc}.
These models also account successfully for the centrality dependence of 
the suppression pattern~\cite{Wang:2003aw,Drees:2003zh,Dainese:2004te}, 
and the reduction of leading back-to-back 
correlation~\cite{Wang:2003aw,Drees:2003zh,Dainese:2004te}. However, 
they tend to underpredict~\cite{Drees:2003zh,Dainese:2004te} the elliptic 
flow $v_2(p_T)$ at high transverse momentum which is thought to originate 
from parton energy loss in an azimuthally asymmetric 
geometry~\cite{Wang:2000fq,Gyulassy:2000gk}. 

In nucleus-nucleus collisions at RHIC and at the CERN SPS, there is 
strong experimental evidence that the produced medium is - if at all - 
only {\it locally} equilibrated and is thus characterized only {\it locally} 
by its energy density. Measurements of low-$p_T$ inclusive hadron 
spectra~\cite{Adler:2003cb,Adams:2003xp} and their
azimuthal asymmetry~\cite{Ackermann:2000tr,Adcox:2002ms,Adler:2003kt} 
support the picture that different hadron species emerge from a common 
medium which has built up a strong collective velocity 
field~\cite{Schnedermann:1993ws,Huovinen:2001cy,Kolb:2001qz,Teaney:2000cw,Kolb:2003dz}. 
These measurements are broadly consistent with calculations based on ideal 
hydrodynamics~\cite{Kolb:2003dz,Kolb:2001qz,Teaney:2000cw,Hirano:2002ds,Retiere:2003kf}, 
in which the dynamic behavior of 
the produced QCD matter is fully specified by its equation of state 
$p = p(\epsilon,T,\mu_B)$ which enters the energy momentum tensor
\begin{equation}
  T^{\mu\nu}(x) = \left(\epsilon + p \right)\, u^{\mu}\, u^{\nu}\, 
                  - p\, g^{\mu \nu}\, . 
  \label{1.2} 
\end{equation}
For the case of a longitudinal Bjorken-type flow field
$u^{\mu} = \left(1,\vec{\beta}\right)/\sqrt{1 - \beta^2}$, 
$\vec{\beta} = \beta\, \hat{z}$, the 
longitudinal component of the energy momentum tensor increases 
from $T^{zz} = p$ to $T^{zz} = p + \Delta p$, where
$\Delta p = (\epsilon + p) u^z\, u^z = 4\, p\, \beta^2/(1-\beta^2)$ 
for the equation of state of an ideal gas, $\epsilon = 3\, p$. 
For a rapidity difference $\eta = 0.5, 1.0, 1.5$ between the rest 
frame which is longitudinally comoving with the jet, and the rest
frame of the medium, the component $T^{zz}$ of the energy-momentum 
tensor ``seen'' by the hard parton is thus increased by 
a factor $1, 5, 18$, respectively. It is thus reasonable to assume 
that the momentum transfer from the medium to a test particle such 
as a hard parton does not depend solely on the local energy density 
$\epsilon$, but rather on the energy momentum tensor (\ref{1.2})
which involves a significant directed collective flow field 
$u_{\mu}(x)$~\cite{Armesto:2004pt}. 

\begin{figure}[h]
\begin{center}
\includegraphics[width=7.3cm,angle=-90]{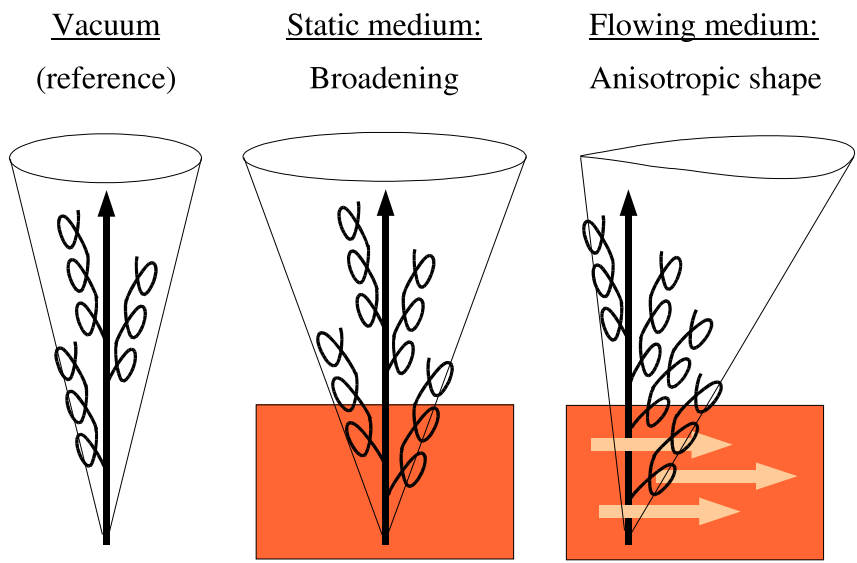}
\end{center}
\vspace{0.5cm}
\caption{Sketch of the expected energy or multiplicity distribution 
of a jet fragmenting i) in the vacuum (left), ii) in a medium 
which is longitudinally comoving with the rest frame of the jet 
(middle) and 
iii) in a medium which is longitudinally boosted with respect to the
rest frame of the jet (right).}
\label{fig1}
\end{figure}
%

Figure~\ref{fig1} sketches the qualitative picture first advocated
in Ref.~\cite{Armesto:2004pt}:
A jet which fragments inside a medium is known to broaden its
shape and to soften its multiplicity distribution. But if the
medium exhibits a collective motion, then a smaller local energy
density is sufficient for the same net momentum transfer to the
hard parton and thus for the same medium-induced parton energy  
loss. Moreover, the directed momentum transfer can be expected 
to break the rotational symmetry of the jet shape in the 
$\eta \times \phi$-plane. In this work, we give a detailed
description of the formalism incorporating these effects and 
we explore observable consequences of the resulting 
interplay of oriented and random momentum transfer to a hard parton.

For each jet, rotational symmetry in the $\eta \times \phi$-plane 
is broken even in the absence of a medium, mainly for
two reasons: Statistically, any finite multiplicity 
distribution of a rotationally symmetric sample breaks the 
rotational symmetry. 
If this were the only source of symmetry breaking, one could 
search for medium-induced asymmetries in realigned jet samples, 
similar to the analysis of elliptic flow in realigned event 
samples~\cite{Ollitrault:bk,Borghini:2000sa}. In addition, however,
the $k_T$-ordering of the DGLAP parton shower implies 
that the first parton splitting in the shower contains significantly 
more transverse momentum than the second, thus leading to a dynamical 
asymmetry in the $\eta \times \phi$-plane. Both
effects lead to a symmetry breaking in a {\it random} direction in 
the $\eta \times \phi$-plane - thus rotational symmetry is restored
in sufficiently large jet samples. To search for symmetry breaking 
effects caused by collective motion in 
$\eta \times \phi$-distributions of jet energy and jet multiplicity, 
it is thus important 
to control experimentally the direction of this collective motion.
Based on these arguments, we foresee two classes of applications 
for our calculations:

\begin{figure}[h]\epsfxsize=13.7cm
\centerline{\epsfbox{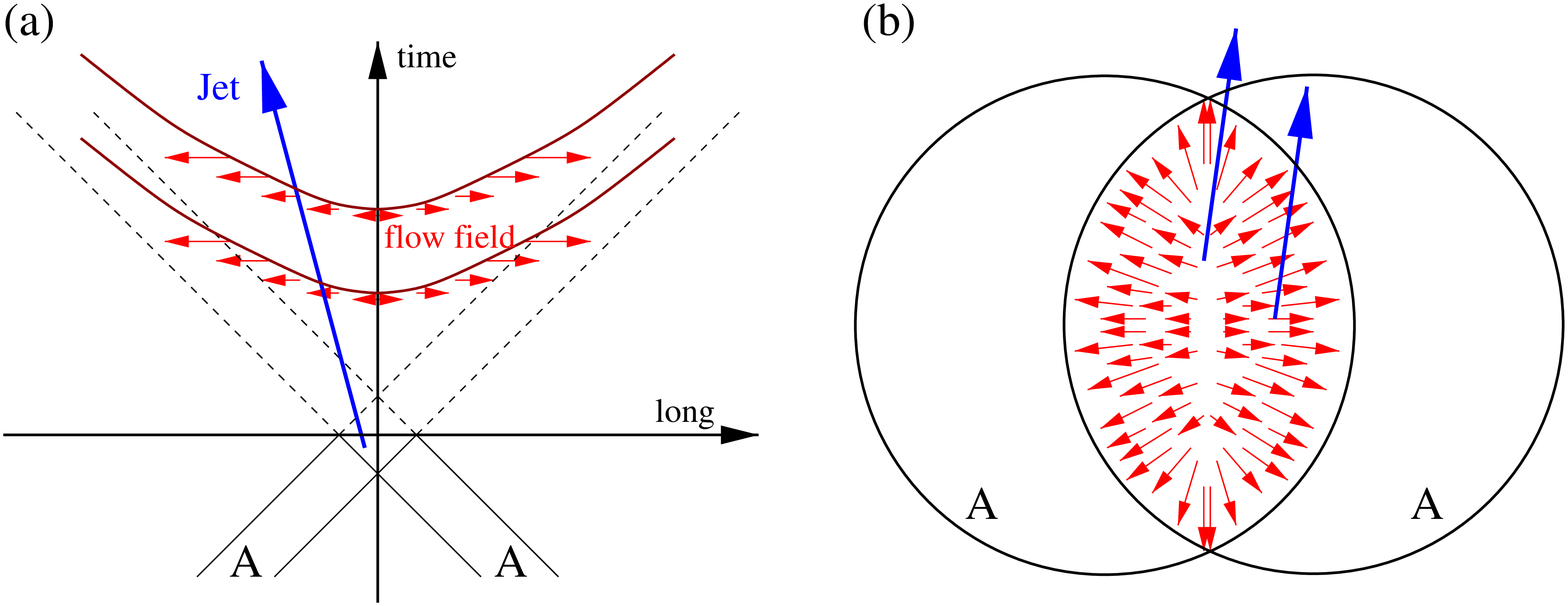}}
\vspace{0.5cm}
\caption{Schematic view of two scenarios in which jets interact
with collective flow fields: a) If the hard parton is not produced
in the Lorentz frame longitudinally comoving with the medium, or
if the longitudinal collective flow does not show Bjorken scaling,
then the parton interacts with a flow component parallel to the
beam. b) On its propagation in the transverse direction, 
hard partons generically test transverse flow components,
except for the special trajectories which are parallel to the
flow field.}
\label{fig2}
\end{figure}
%

First, in general, a hard parton needs not be produced in the Lorentz 
frame which is longitudinally comoving with the medium; and even if it 
is produced in the longitudinally comoving frame, it will in general 
not stay in this frame during the entire time evolution of the medium.
This is so since the hard parton moves -- like any effectively massless
particle -- on a straight light-like line in the $(z,t)$-diagram,
whereas the collective flow field is expected to show significant 
deviations~\cite{Hirano:2003yp,Bearden:2004yx} from Bjorken expansion 
and will thus intersect this straight 
line. In such cases, the collective component of the momentum transfer 
to the hard parton is directed along the beam axis. Hence, averaged 
samples of medium-modified jet shapes and jet multiplicities can be 
expected to show an asymmetry which is preferentially oriented along 
the beam direction in the $\eta \times \phi$-plane. [At mid-rapidity,
the jet sample must be symmetric with respect to the $\eta \to -\eta$
mirror symmetry, but -- in general -- it will not be rotationally
symmetric in the $\eta \times \phi$-plane.] In the case of a 
significant transverse collective flow, the analogous argument implies
the occurrence of jet asymmetries preferentially oriented along the 
$\phi$-direction. In general, the
size and orientation of the jet asymmetry depends on how hard
processes are spatially distributed in the dynamical expansion
scenario. 

Second, flow effects manifest themselves
not only in the azimuthal asymmetries of jet observables, but also
in inclusive high-$p_T$ hadron spectra. In the presence
of a flow field, a smaller local energy density is sufficient
for the same net momentum transfer to the hard parton and thus
for the same medium-induced parton energy loss. This is relevant for
the interpretation of the nuclear modification factor
in terms of a local energy density. Moreover, hard partons propagating 
parallel to a flow field 
can be expected to suffer less momentum transfer and hence less 
parton energy loss than those traveling along non-parallel 
trajectories. In this way, low-$p_T$ collective flow can induce
high-$p_T$ azimuthal asymmetry by selective elimination of those hard 
partons which propagate with significant inclination against the flow 
field. We show that this can yield to a sizeable additional contribution
to high-$p_T$ $v_2$.

This paper is organized as follows: In section~\ref{sec2}, we introduce
the formalism in which the effects of anisotropic momentum transfer
on parton energy loss are calculated. In section~\ref{sec3}, we calculate
the induced asymmetry of the medium-dependent gluon energy distribution
and in section~\ref{sec4} we analyze the resulting anisotropic jet
energy distribution. Up to this point, the medium will be characterized
schematically by the momentum scale $\mu$ which determines the random
momentum transfer to the hard parton, and the vector ${\bf q}_0$
which specifies the oriented momentum transfer. The ratio $q_0/\mu$
indicates the relative strength of collective flow. In section~\ref{sec5},
we then discuss how to embed this calculation in a dynamical expansion
scenario and we estimate the flow-induced parton energy loss contribution
to high-$p_T$ $v_2$. Finally, we summarize the main conclusions.

\section{The formalism}
\label{sec2}
The starting point of our calculation is the energy distribution of
gluons into which the initially produced parent parton fragments, 
\begin{equation}
  \omega\frac{dI^{\rm tot}}{d\omega\, d{\bf k}}
  = \frac{E_T - \Delta E_T}{E_T} 
    \omega\frac{dI^{\rm vac}}{d\omega\, d{\bf k}}
  + \omega\frac{dI^{\rm med}}{d\omega\, d{\bf k}}\, .
  \label{2.1}
\end{equation}
Here, $E_T$ denotes the total energy of the hard parton
and $\Delta E_T = \int d\omega\, d{\bf k}\,
\omega\, \frac{dI^{\rm med}}{d\omega\, d{\bf k}}$ is that
part of the total energy which is redistributed by 
medium-induced radiation. The factor 
$\left(E_T - \Delta E_T \right) / E_T$
ensures energy-momentum conservation.

In the absence of a nuclear environment, the parton fragments
according to the distribution $I^{\rm tot} = I^{\rm vac}$. 
In the medium, the parent parton radiates additional gluons 
due to medium-induced multiple scattering. This medium-induced
gluon radiation has been calculated to leading order in $1/E$, resumming 
an arbitrary number of scattering centers~\cite{Baier:1996sk,Zakharov:1997uu,Wiedemann:2000za,Gyulassy:2000er}. It depends to leading order 
in $1/E$ on the in-medium pathlength $L$ and on the average squared 
transverse momentum transferred to the hard parton per unit pathlength. 
The latter property of the medium is parameterized differently in
different approaches, e.g. in terms of the BDMPS transport 
coefficient $\hat{q}$~\cite{Baier:1996sk,Zakharov:1997uu,Wiedemann:2000za}, 
or as the product of the longitudinal density 
of scattering centers $n_0$ along the parton trajectory, and their 
typical momentum transfer $\mu^2$~\cite{Wiedemann:2000za,Gyulassy:2000er}. 
These parameterizations are known 
to lead to equivalent results for the medium-induced gluon 
radiation~\cite{Salgado:2003gb}. In what follows, we use
the single-hard scattering approximation, in which 
the effects of multiple scattering are characterized by
the number of effective scattering centers $n_0\, L$
times the radiation off a single scattering center.
The elastic scattering cross section is modeled in terms of
Debye-screened Yukawa-potentials
\begin{equation}
 \vert a({\bf q})\vert^2 = \frac{\mu^2}{\pi (\mu^2 + {\bf q}^2)^2}\, .
  \label{2.2}
\end{equation}
The medium-induced gluon radiation is given 
by~\cite{Wiedemann:2000za,Gyulassy:2000er,Salgado:2003gb} 
\begin{equation}
 \omega \frac{dI^{\rm med}}{d\omega\, d{\bf k}} = \frac{\alpha_s}{(2\pi)^2}
 \frac{4\, C_R\, n_0}{\omega} \, 
 \int d{\bf q}\, \vert a({\bf q})\vert^2\, 
 \frac{ {\bf k}\cdot {\bf q}}{{\bf k}^2}\, 
 \frac{-L \frac{({\bf k} + {\bf q})^2}{2\omega} + 
       \sin\left(   L \frac{({\bf k} + {\bf q})^2}{2\omega}\right)}
     {  \left[({\bf k} + {\bf q})^2 / 2\omega\right]^2} \, .
 \label{2.3}
\end{equation}
This radiation spectrum is for a time-independent homogeneous
medium. However, it also applies to a time-dependent density of
scattering centers if the density $n_0$ is replaced by an
appropriate time-average (see Ref.~\cite{Salgado:2002cd,Salgado:2003gb}
and section~\ref{sec5a} below).

If a flow component is directed orthogonally to the parton trajectory,
then the momentum transfer from the medium is anisotropic. We denote
by $\vec{q}_0 = ({\bf q}_0,q_0^l)$ the directed momentum transfer to the
hard parton which is parallel to the spatial components of the flow 
field $u_{\mu}(x)$. Here and in the following, 
transverse vectors lie in the plane orthogonal to the trajectory 
of the hard parton while longitudinal components are parallel to
this trajectory. In the high-energy limit, momentum transfers 
parallel to the hard parton are negligible. Thus, the effect of 
collective flow on the medium-induced radiation (\ref{2.3})
can be accounted for by using an anisotropic scattering potential  
\begin{equation}
  \vert a({\bf q})\vert^2 = \frac{\mu^2}{\pi 
  \left[ \left({\bf q} - {\bf q}_0\right)^2 + \mu^2\right]^2}\, .
  \label{2.4}
\end{equation}
The parameters $\mu$ and $\vert {\bf q}_0\vert$ characterize
the strength of the random and directed momentum transfers
from the medium to the hard test particle, respectively. 
The component $q_0^l$ is parallel 
to the parton trajectory and does not enter our calculation. 
We work in radial coordinates, 
\begin{equation}
  d{\bf q} = q\, dq\, d\varphi\, , \qquad 
  d{\bf k} = k\, dk\, d\alpha\, , 
  \label{2.5}
\end{equation} 
where $\alpha$ denotes
the angle between the transverse momenta ${\bf k}$ and ${\bf q}_0$.
The $\varphi$-integration in (\ref{2.3}) can then be done analytically.
We always work in the frame longitudinally comoving with the hard 
parton in which the parton propagates orthogonal to the beam direction.

%
\begin{figure}[h]\epsfxsize=10.7cm
\centerline{\epsfbox{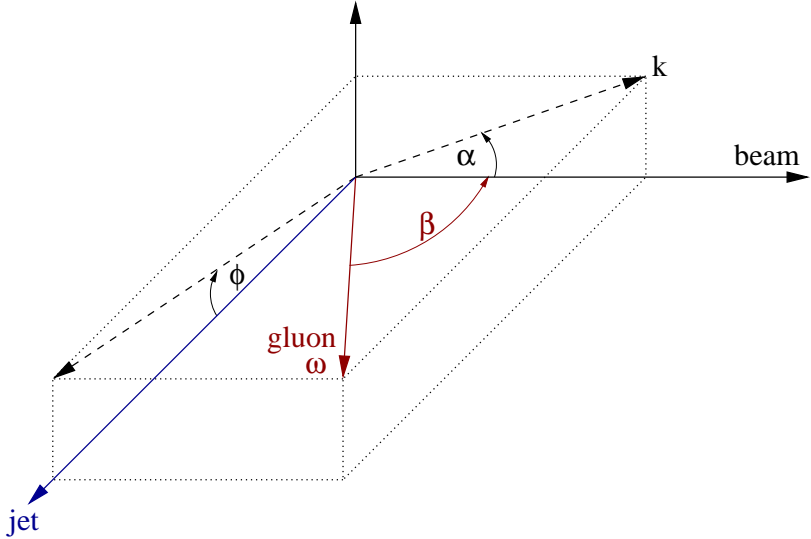}}
\vspace{0.5cm}
\caption{Definition of kinematic variables of a gluon emitted
inside a jet cone. Variables are defined in the Lorentz frame 
which is longitudinally comoving with the jet.
}\label{fig3}
\end{figure}
%

We now consider more explicitly the case of a longitudinal flow 
component parallel to the beam direction, see Fig.~\ref{fig3}.
To express the radiation spectrum as a function of
pseudorapidity $\eta$ and azimuthal angle $\phi$ with respect
to the center of the jet at $\eta = \phi = 0$, we write
the pseudorapidity of an emitted gluon as 
$\eta = - \ln \tan \frac{\beta}{2}$ where $\beta$ is the
angle between the gluon momentum and the beam axis. We have
\begin{equation}
  \tan\beta = \frac{\omega}{k\, \cos\alpha}
  \, \sqrt{1 - \left(\frac{k}{\omega}\right)^2 \cos^2\alpha }
  \, ,\qquad
  \tan\phi = \frac{k\, \sin\alpha}{\sqrt{\omega^2 - k^2}}\, .
  \label{2.6}
\end{equation}
Inversion leads to
\begin{eqnarray}
  \cos\alpha = \frac{ \sinh \eta}{\sqrt{ \cosh^2\eta - \cos^2\phi}}\, ,
  \qquad
  \frac{k}{\omega} = \frac{\sqrt{\cosh^2\eta - \cos^2\phi}}{\cosh \eta}\, .
  \label{2.7}
\end{eqnarray}
The Jacobian for the transformation to jet observables $\eta$, $\phi$
reads 
\begin{equation}
  k\, dk\, d\alpha = \omega^2\, \frac{\cos\phi}{\cosh^3\eta}\, 
  d\eta\, d\phi\, .
  \label{2.8}
\end{equation}
Our final expression is 
\begin{eqnarray}
 \omega \frac{dI^{\rm med}}{d\omega\, d\eta\, d\phi} &=& 
 \omega^3\, \frac{\cos\phi}{\cosh^3\eta}\, 
 \frac{\alpha_s\, C_R}{\pi^2}
 \frac{4\, n_0\, \mu^2}{{\bf k}^2} \, 
 \int_0^{\infty} d{\bf q}^2\, 
 \frac{\frac{L\, {\bf q}^2}{2\omega} - 
       \sin \frac{L\, {\bf q}^2}{2\omega}}
     { {\bf q}^4}
     \nonumber \\
 && \times \frac{{\bf k}^2\, \left(q^2 + \mu^2 + ({\bf q}_0 + {\bf k})^2\right)
       - 2 {\bf q}^2 {\bf k}\cdot ({\bf q}_0 + {\bf k})}
 {\left[ \left({\bf q}^2 + \mu^2 
         - \left({\bf q}_0 + {\bf k}\right)^2 \right)^2 
         + 4 \mu^2 \left({\bf q}_0 + {\bf k}\right)^2 \right]^{3/2}}\, 
 \, ,
 \label{2.9}
\end{eqnarray}
where ${\bf k}$ and the angle $\alpha$ are given by (\ref{2.7})
in terms of $\eta$ and $\phi$. The case of a transverse flow 
component is obtained by obvious rotations and redefinitions of
the vectors in Fig.~\ref{fig3}.

\section{Transverse momentum dependence of the medium-induced gluon 
radiation in the presence of flow}
\label{sec3}

In this section, we discuss the generic properties of the medium-induced 
gluon radiation (\ref{2.3}) in the presence of collective flow.
To this end, it is convenient to change to 
dimensionless variables
\begin{eqnarray}
  \bar\kappa = {\vert{\bf k}\vert}/\mu\, ,\qquad 
  \bar{q} = {\vert{\bf q}\vert}/\mu\, ,\qquad
  \bar\gamma = \bar{\omega}_c/\omega\, ,\qquad 
  \bar{\omega}_c  = \frac{1}{2} \mu^2\, L\, .
  \label{3.1}
\end{eqnarray}
We transform (\ref{2.3}) to radial coordinates
$d{\bf q} = \mu^2 \bar{q}\, d\bar{q}\, d\varphi$
and $d{\bf k} = \mu^2 \bar{\kappa}\, d\bar{\kappa}\, d\alpha$, 
where $\alpha$ denotes
the angle between the transverse momenta ${\bf k}$ and ${\bf q}_0$.
Doing the $\varphi$-integration in (\ref{2.3}), we find
\begin{eqnarray}
 &&\omega \frac{dI^{\rm med}}{d\omega\, \bar{\kappa}\, d\bar{\kappa}\, d\alpha} 
  = \frac{\alpha_s\, C_R}{\pi^2} 2\, n_0\, L\, 
 \int d\bar{q}^2\,
 \frac{ \bar{q}^2 - \frac{1}{\bar{\gamma}}  
       \sin\bar{\gamma} \bar{q}^2 }{\bar{\kappa}^2\,  \bar{q}^4} 
     \nonumber \\
 && \times
 \frac{ \bar{\kappa}^2 \left( \bar{q}^2 + 1 
        + (\bar{q}_0^2 + \bar{\kappa}^2 
               + 2 \bar{q}_0 \bar{\kappa} \cos\alpha ) \right)
        - 2  \bar{q}^2 \left(\bar{\kappa}^2 + \bar{\kappa}\bar{q}_0\cos\alpha \right)}
        {\left[ \left(\bar{q}^2 + 1 - (\bar{q}_0^2 + \bar{\kappa}^2 
               + 2 \bar{q}_0 \bar{\kappa} \cos\alpha )   
                 \right)^2    
                + 4 (\bar{q}_0^2 + \bar{\kappa}^2 
               + 2 \bar{q}_0 \bar{\kappa} \cos\alpha ) \right]^{3/2}}\, .
 \label{3.2}
\end{eqnarray}
%
\begin{figure}[h]\epsfxsize=13.7cm
\centerline{\epsfbox{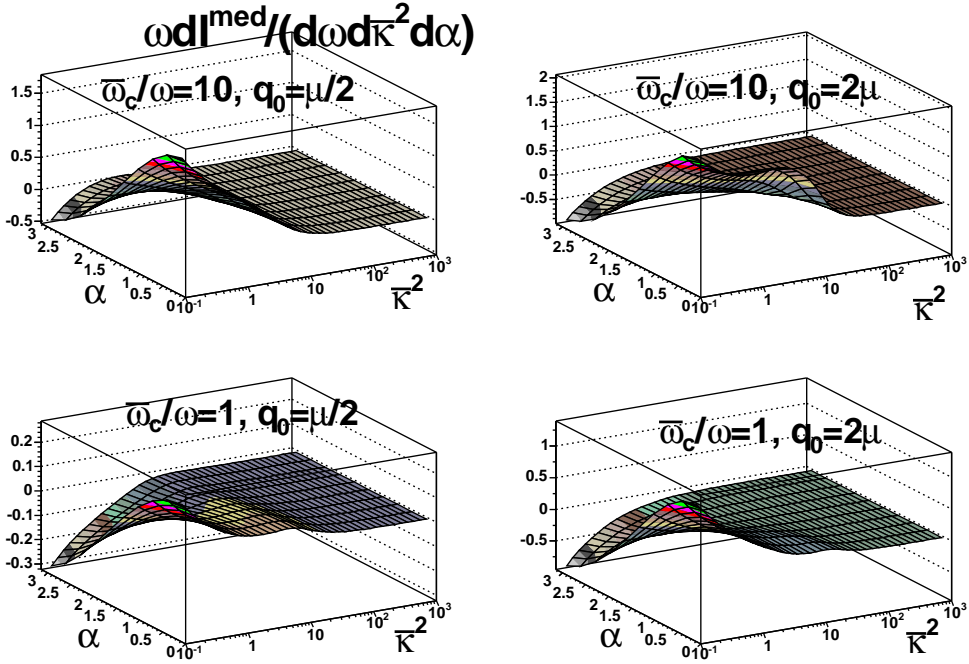}}
\vspace{0.5cm}
\caption{Plot of the medium-induced gluon energy distribution
(\protect\ref{3.2}) as a function of the angle $\alpha$ and
rescaled transverse momentum $\bar{\kappa}^2 = {\bf k}^2/\mu^2$ for 
different values of the rescaled gluon energy 
$\frac{1}{2} \mu^2\, L/ \omega$
and the collective flow strength $q_0$. Plots are for $\mu = 2$ 
GeV and $L = 6$ fm.
}\label{fig4}
\end{figure}
%
In Fig.~\ref{fig4}, we plot the medium-induced
energy distribution as a function of $\bar{\kappa}^2$ and $\alpha$ 
for fixed ratios of $\bar{\gamma} = \bar\omega_c/\omega$. 
Here, $\alpha = 0$ denotes the direction of the collective flow 
vector ${\bf q}_0$. The figure~\ref{fig4} shows clearly that more
energy is deposited in the direction of the collective flow vector 
${\bf q}_0$.
For the same reason, the energy distribution is depleted in the
direction opposite to ${\bf q}_0$, i.e. for $\alpha \sim \pi$. For 
the medium-induced component plotted in Fig.~\ref{fig4}, this shows 
up as a negative contribution, while the total energy distribution 
(\ref{2.1}) stays, of course, positive. 

For non-zero values of the flow vector ${\bf q}_0$, the triple 
differential gluon distribution (\ref{3.2}) has a singular
behavior for $\bar{\kappa} \to 0$,
\begin{equation}
  \lim_{\bar{\kappa} \to 0}\,
  \omega \frac{dI^{\rm med}}{d\omega\, \bar{\kappa}\, 
    d\bar{\kappa}\, d\alpha}
    =  \left\{ \begin{array} 
                  {r@{\qquad  \hbox{for}\quad}l}                 
                  + \infty
                  & - \frac{\pi}{2} < \alpha < \frac{\pi}{2}  \, , \\ 
                  - \infty
                  & \frac{\pi}{2} < \alpha < \frac{3 \pi}{2} \, . 
                  \end{array} \right.
 \label{3.3}               
\end{equation}
Figure~\ref{fig4} displays only finite values of $\bar{\kappa}$,
but the limit (\ref{3.3}) is consistent with the small-$\bar{\kappa}$
behavior seen in Fig.~\ref{fig4}.
This singularity is unphysical. It stems from the fact that the formalism 
leading to (\ref{2.3}) calculates medium-modifications to a perturbative 
parton splitting $\propto \frac{1}{{\bf k}^2}$ without regularization 
of this collinear divergence. At $\bar{\kappa} = 0$, the anisotropic flow
field shifts part of this singularity as a positive contribution
to the half plane $ - \frac{\pi}{2} < \alpha < \frac{\pi}{2}$ 
while depleting the region $ \frac{\pi}{2} < \alpha < \frac{3 \pi}{2}$.
In the $\eta \times \phi$-plane, (\ref{2.9}) integrated over energy 
shows the same singular structure, namely a positive divergence for 
$\eta \to 0^+$ and a negative divergence for $\eta \to 0^-$. 
These two divergences cancel each other if integrated
over an arbitrary small neighborhood around $\bar{\kappa} = 0$ 
(i.e. around $\eta = \phi = 0$). They represent a very small contribution 
to the total jet energy (see the discussion below). 
%
\begin{figure}[h]\epsfxsize=13.7cm
\centerline{\epsfbox{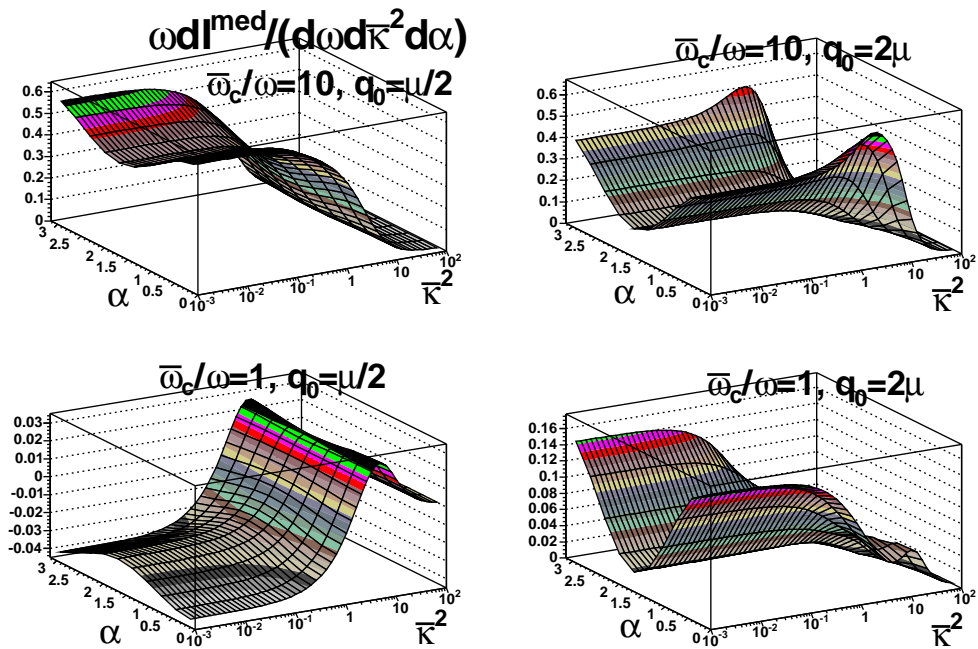}}
\vspace{0.5cm}
\caption{The gluon energy distribution (\ref{3.2}) for the same
parameters as in Fig.~\ref{fig4}, but averaged over the cases
${\bf q}_0$ and $-{\bf q}_0$.
}\label{fig5}
\end{figure}
%

In Fig.~\ref{fig5}, we plot the gluon energy distribution (\ref{3.2})
for a sample of medium-modified jet shapes for which the orientation
$\pm {\bf q}_0$ of the collective flow is known but the directions
${\bf q}_0$ and $-{\bf q}_0$ are equally likely. 
The figure clearly indicates that the singular behavior for
$\bar{\kappa} \to 0$ disappears. For fixed values of 
$\bar{\gamma} = \frac{\bar\omega_c}{\omega}$, the energy distribution has
a finite constant value for $\bar{\kappa} \to 0$. By construction, 
the obtained distribution is
symmetric around $\alpha = \frac{\pi}{2}$, but it shows a marked
angular dependence. 

\section{Collective flow leads to anisotropic jet energy distributions}
\label{sec4}

In this section, we discuss the medium-modification of jet energy
distributions. Our starting point is the gluon energy distribution
(\ref{2.1}). We constrain the vacuum contribution $I^{\rm vac}$ 
of this spectrum by data on the energy fraction of 
a jet contained in a subcone of radius 
$R = \sqrt{\eta^2 + \phi^2}$,
\begin{eqnarray}
  \rho_{\rm vac}(R) &\equiv& \frac{1}{N_{\rm jets}} \sum_{\rm jets}
  \frac{E_T(R)}{E_T(R=1)}
  \nonumber\\
  &=& 1 -  \frac{1}{E_T} \int d\omega \int^\omega d{\bf k}\, 
  \Theta\left(\frac{k}{\omega} - R\right)\, 
  \omega\frac{dI^{\rm vac}}{d\omega\, d{\bf k}}\, .
  \label{4.1}
\end{eqnarray}
For the jet shape 
$\rho_{\rm vac}(R)$, we use the parameterization~\cite{Abbott:1997fc} 
of the Fermilab $D0$ Collaboration determined for jets in the range 
$\approx 50 < E_t < 170$ GeV and opening cones $0.1 < R < 1.0$, 
\begin{equation}
   \rho_{\rm vac}^{(D0)}(R) = A\, R^{0.1}+ B\, R^{0.3} + C\, R^{0.5} 
                              + D\, R^{0.7} + E\, R^{0.9}\, ,
                              \label{4.2}
\end{equation}
where
\begin{eqnarray}
  A(E_T) &=& -3.47 + 0.85\cdot 10^{-2} E_T - 0.25\cdot 10^{-4} E_T^2\, ,
  \nonumber\\
    D(E_T) &=&  3.30 - 0.77\cdot 10^{-2} E_T + 0.22\cdot 10^{-4} E_T^2\, ,
   \nonumber\\
    B &=& 9.75\, ,\qquad  C = -8.32\, ,\qquad  E = -0.30\, .
    \label{4.3}
\end{eqnarray}
At very small values of $R$, this parameterization turns negative. 
We cure this unphysical behavior by matching (\ref{4.2}) with 
a third order polynomial which smoothly interpolates to 
$\rho_{\rm vac}(R=0) = 0$,
\begin{equation}
 \rho_{\rm vac}(R) = 
 \left\{ \begin{array} 
                  {r@{\qquad  \hbox{for}\quad}l}
                  \rho_{\rm vac}^{(D0)}(R)
                  & R > 0.04 \, ,\\ 
                  a R^2 + b R^3
                  & R < 0.04\, .  
                  \end{array} \right.
                \label{4.4}
\end{equation}
The parameters $a$ and $b$ in (\ref{4.4}) are fixed by
requiring that $\rho_{\rm vac}(R)$ and its first derivative
are continuous.

The medium-induced part of the jet energy distribution is
determined from (\ref{2.9}), 
\begin{equation}
  \frac{dE^{\rm med}}{d\eta\, d\phi}
  = \int_0^E d\omega\, 
  \omega\, \frac{dI^{\rm med}}{d\omega\, d\eta\, d\phi}\, .
  \label{4.5}
\end{equation}
It contains a fraction $\frac{\Delta E_T}{E_T}$ of the available jet 
energy, 
\begin{equation}
  \Delta E_T = \int d\eta \int d\phi\, 
  \frac{dE^{\rm med}}{d\eta\, d\phi}\, .
  \label{4.6}
\end{equation}
For the vacuum contribution, the corresponding distribution 
is defined by the vacuum jet shape (\ref{4.4}), 
\begin{equation}
  \frac{dE^{\rm vac}}{d\eta\, d\phi} = 
  \left(E_T - \Delta E_T \right)\, 
  \frac{d\rho_{\rm vac}}{2\pi\, R\, dR}\, .\
  \label{4.7}
\end{equation}  
Here, the prefactor $\left(E_T - \Delta E_T \right)$ ensures 
energy conservation.

\subsection{Harmonic expansion of jet energy distribution}

In this section, we characterize the flow-induced asymmetry
of jet energy distributions in terms of a harmonic analysis in the
$\eta \times \phi$-plane. We introduce radial coordinates
in this plane,
\begin{eqnarray}
  R &=& \sqrt{\eta^2 + \phi^2}\, ,
  \label{4.8} \\
  \alpha^\prime &=& \arctan \left( \frac{\phi}{\eta} \right)\, .
  \label{4.9}
\end{eqnarray}
The flow field points in the direction $\alpha' = 0$. We calculate
now the jet energy distribution in the $\eta \times \phi$-plane,
\begin{equation}
  \frac{dE_T}{R\, dR\, d\alpha^\prime}
  =  \frac{dE_T}{d\eta\, d\phi}
  =  \frac{dE_T^{\rm vac}}{d\eta\, d\phi}
   +  \frac{dE_T^{\rm med}}{d\eta\, d\phi}\, .
  \label{4.10}
\end{equation}
In the absence of flow, the radiation spectrum (\ref{3.2}) is rotationally 
symmetric in the coordinates $\bar{\kappa}$ (or ${\bf k}$) and $\alpha$.
However, the energy distribution (\ref{4.10}) is elongated in the
$\phi$-direction due to the Jacobian (\ref{2.8}) in the coordinate
transform. In general, this reduces the effect of $\eta$-broadening 
due to longitudinal flow, and it can be corrected for analytically.
However, this asymmetry is rather small ($< 10 \%$) for small cone 
sizes ($R < 0.3$), and can be neglected safely in the following 
discussion. 
To analyze the asymmetries of the jet energy distribution
(\ref{4.10}) in the $\eta \times \phi$-plane, we use
a harmonic expansion
\begin{equation}
  \frac{dE_T}{R\, dR\, d\alpha^\prime} =
  E^{(0)}(R) + 2 \sum_{n=1}^\infty  E^{(n)}(R)\, 
  \cos \left(n \alpha^\prime \right)\, .
  \label{4.11}
\end{equation}
The coefficients proportional to $\sin \left(n \alpha^\prime \right)$
cancel since the jet energy distribution (\ref{4.10}) is by construction
symmetric with respect to $\alpha^\prime \to - \alpha^\prime$. 
\vspace{-0.5cm}
\begin{figure}[h]\epsfxsize=14.7cm
\centerline{\epsfbox{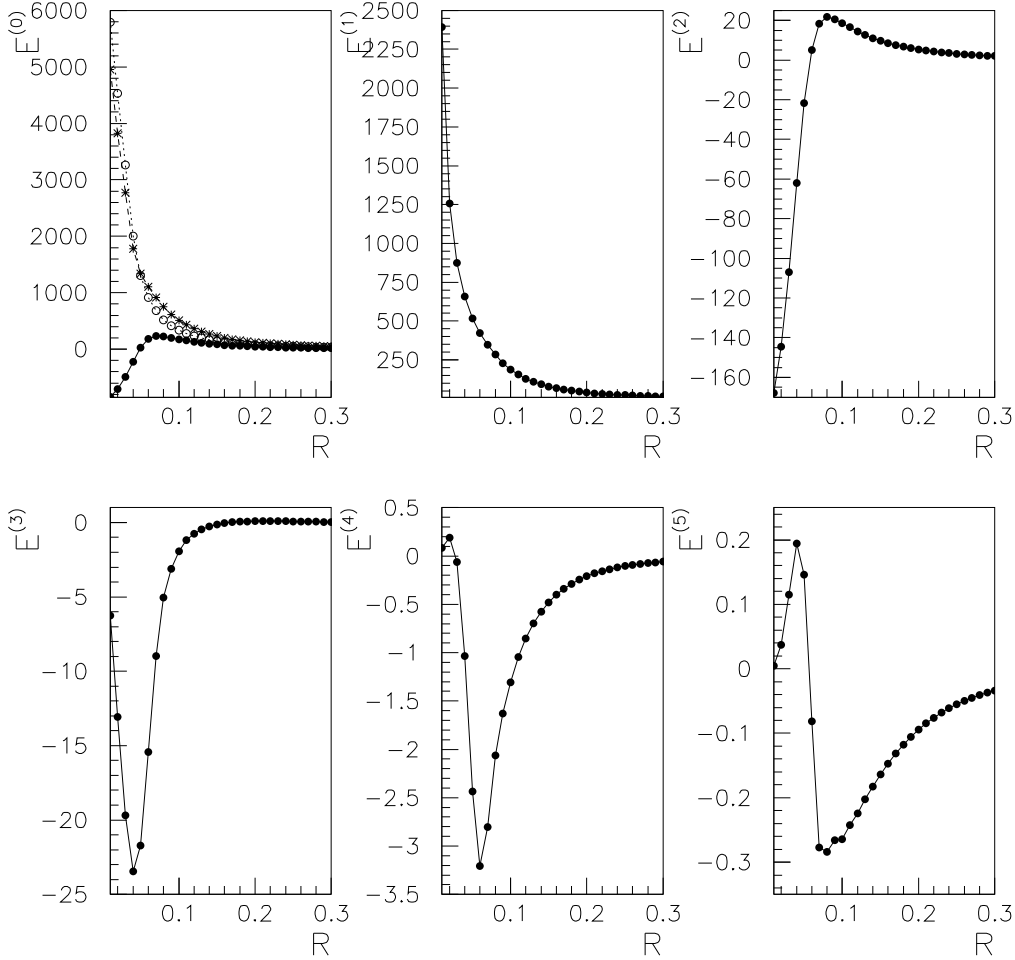}}
\vspace{0.5cm}
\caption{ The harmonic coefficients which characterize the asymmetry
of the jet energy distribution of a $100$ GeV jet in the 
$\eta \times \phi$-plane. Parameters given in the text.
}\label{fig6}
\end{figure}

For illustration, we now study the case of a jet of total energy
$E_T = 100$ GeV traversing a medium in which 
$\Delta E_T = 23$ GeV were redistributed in phase space due
to medium effects. This parton energy loss is obtained e.g. for 
an in-medium pathlength of nuclear size ($L = 6$ fm), a momentum 
transfer per scattering center of $\mu = 1$ GeV and a collective flow
effect of the same size $q_0 = \mu$, with an effective coupling 
constant in (\ref{2.3}) fixed to $n_0\, L\, \alpha_s C_R = 1$. 
This corresponds 
to a gluon jet ($C_R = C_A = 3$) with a reasonable perturbative
coupling $\alpha_s = 1/3$ and an opacity $n_0\, L = 1$. 
Changes in the coupling can be absorbed in a redefinition of the
density of scattering centers. Recent model comparisons with RHIC
data support the picture of an opaque medium which may result 
in significantly larger medium-induced modifications than the
ones modeled here. However, to illustrate the sensitivity
of jet shape measurements, we prefer to work with a sizeable but
relatively small effect.

In Fig.~\ref{fig6}, we show the first 6 harmonic coefficients
extracted for the jet energy distribution (\ref{4.11}).
Since the vacuum term (\ref{4.4}) is rotationally symmetric, 
it contributes only
to $E^{(0)}(R)$. The total jet energy is obtained by integrating 
this zeroth moment over $R$. The shape of the
medium-induced part of $E^{(0)}(R)$ is negative at small $R$
which indicates the medium-induced depletion of the jet energy
in this region of phase space. This is the result of multiple 
scattering which broadens the jet energy distribution by moving a 
fraction of the total jet energy to larger values of $R$. 
The higher moments $E^{(n)}(R)$,
$n \geq 1$, contain information about asymmetries in the energy
distribution but do not contribute to the total jet energy. 
For the case of averaged jet samples over opposite flow directions 
${\bf q}_0$ and $-{\bf q}_0$, the odd harmonic moments vanish
while the even ones stay the same. The absolute size of the
harmonic moments decreases by approximately one order of magnitude
per moment $n$. This indicates that the first and second moments
are sufficient to characterize the asymmetries of the jet shape.  
We note that to reconstruct from the moments $E^{(0)}(R)$, 
$E^{(1)}(R)$, $E^{(2)}(R)$ the jet shape, one has to integrate 
over $R\, dR$ - this tames significantly the large values of these
moments in the region $R \sim 0$. We expect that most of the 
experimentally accessible structures lies in the range 
$0.05 < R < 0.3$.

The technical advantage of the harmonic expansion (\ref{4.11}) is
that the unphysical singularities of the vacuum contribution 
(\ref{4.7}) and the medium contribution (\ref{4.5}) at $R= 0$ 
can be removed easily. In the harmonic coefficients $E^{(n)}(R)$ 
of (\ref{4.11}), these singularities appear in the odd moments at 
$R \to 0$. The smallest value calculated for Fig.~\ref{fig6}
is for $R=0.01$. For much smaller values of $R$, we find numerically 
$E^{(1)}(R=10^{-4}) \sim 2\,\cdot 10^{5}$ and 
$E^{(1)}(R=10^{-6}) \sim 2\,\cdot 10^{7}$. For the figures presented
in this work, we cut off this artificial small-$R$ structure in the 
harmonics and plot the jet energy distribution according to 
(\ref{4.11}). As explained above, the presence of these singularities
indicates that our formalism becomes unreliable in the collinear 
region of very small $R$. For the arguments in this paper, this is 
not a problem since only a small amount ($< 5 \%$ for $R < 0.01$) 
of the total jet energy lies inside this small phase space region. 

\subsection{Profile and displacement of jet energy distribution}

\begin{figure}[h]\epsfxsize=9.7cm
\centerline{\epsfbox{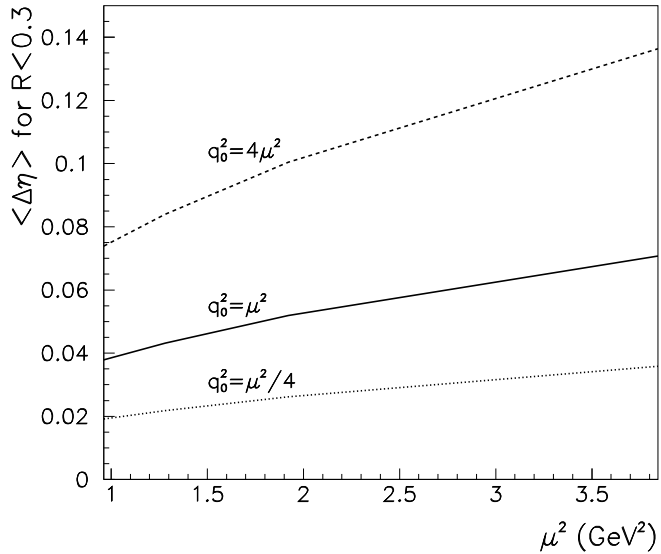}}
\vspace{0.5cm}
\caption{The displacement $\Delta \eta$ of the calorimetric center
of the jet cone as a function of collective flow strength $q_0$ and 
average momentum transfer $\mu$ from the medium. 
}\label{fig7}
\end{figure}

In general, collective flow shifts the calorimetric jet center
and distorts the shape of the jet energy distribution. 
Any calorimetric jet finding algorithm can be expected to center
the cone around the medium-displaced jet center. Fig.\ref{fig7} 
shows this displacement $\Delta \eta$ as a function of the
collective flow strength, calculated by averaging the jet energy 
distribution (\ref{4.10}) over a central cone of size $R < 0.3$. 
For a fixed average momentum transfer $\mu$ per scattering center, 
the displacement grows approximately linearly with the directed
momentum transfer $q_0$. Also, $\Delta \eta$ grows approximately 
linear with the average momentum transfer $\mu$ for a fixed ratio
$q_0/\mu$. The overall size of the displacement is rather small: 
a displacement of size $\Delta \eta = 0.1$ results only for rather 
large parameter values (e.g. $q_0^2 = 4\, \mu^2 = 8$ GeV$^2$)
which correspond to a large medium-induced average energy loss  
($\Delta E_T \approx 62$ GeV). This is consistent with
the picture that the most energetic jet fragments are radiated 
collinear. These energetic components which dominate the energy
distribution, are shifted very little in the $\eta \times \phi$-plane 
even if they pick up a significant transverse momentum. 
%
%

\begin{figure}[h]\epsfxsize=14.7cm
\centerline{\epsfbox{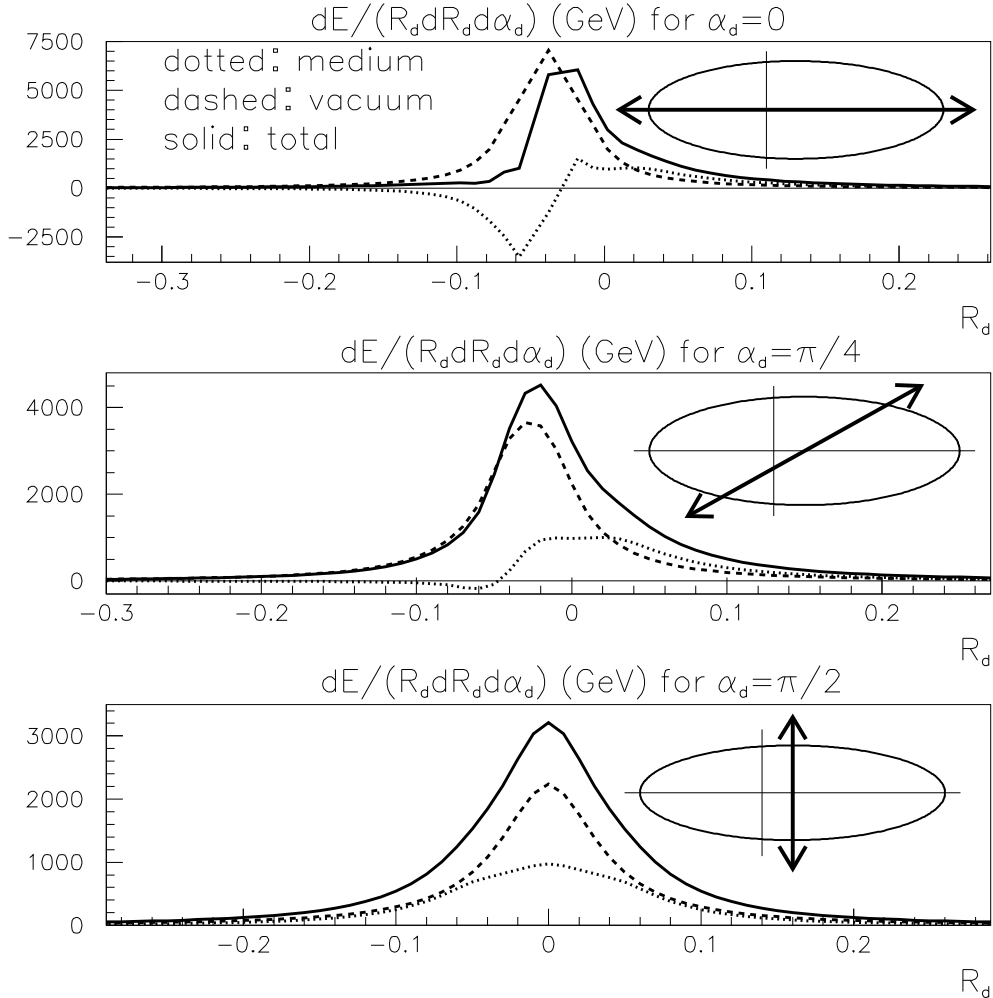}}
\vspace{-0.5cm}
\caption{The jet energy distribution (\ref{4.10}),
plotted along different cuts in the $\eta \times \phi$-
plane, as indicated. The variables $R_d$ and $\alpha_d$ denote 
the distance and orientation 
with respect to the displaced calorimetric center of the 
jet cone. Parameters are the same as those of Fig.\protect\ref{fig6}.
}\label{fig8}
\end{figure}

In Fig.~\ref{fig8}, we show different one-dimensional cuts through 
the jet energy distribution (\ref{4.10}). These cuts go through
the displaced jet center and are quantified by the radial
coordinates $R_d$ and $\alpha_d$ of the displaced center. 
Along the beam direction ($\alpha_d = 0$), the jet energy distribution 
is shifted with the flow field, see Fig.~\ref{fig8}.
The medium-induced part of the 
jet energy distribution takes negative values in the region
of phase space which is depleted due to medium effects. We
observe in particular a pronounced long tail of the distribution
in the direction of the flow. We attribute this tail to the soft 
jet fragments which can be displaced significantly in $\eta$ by a
typical momentum transfer from the medium. In
contrast, in the direction orthogonal to the collective flow, 
one observes a numerically small medium-induced broadening 
of the jet energy distribution, which is not accompanied
by a displacement. 
\vspace{-0.5cm}
%

For measurements at mid-rapidity, one can access experimentally only the
orientation but not the direction of the collective flow
component. Then, each jet of the sample is 
positioned around its calorimetric center which is shifted with 
equal probability in the positive or negative beam direction.
We have given numerical results for this case already in a
previous exploratory study~\cite{Armesto:2004pt}. 
The main conclusion is that the
jet energy distribution can broaden significantly along the
orientation of the flow. The analogous conclusion holds for the
case of a strong transverse collective flow field, where only 
$\phi \to -\phi$-symmetrized samples are measurable.

\subsection{Asymmetries in the jet multiplicity distributions}
\label{sec4c}

\begin{figure}[h]\epsfxsize=14.7cm
\centerline{\epsfbox{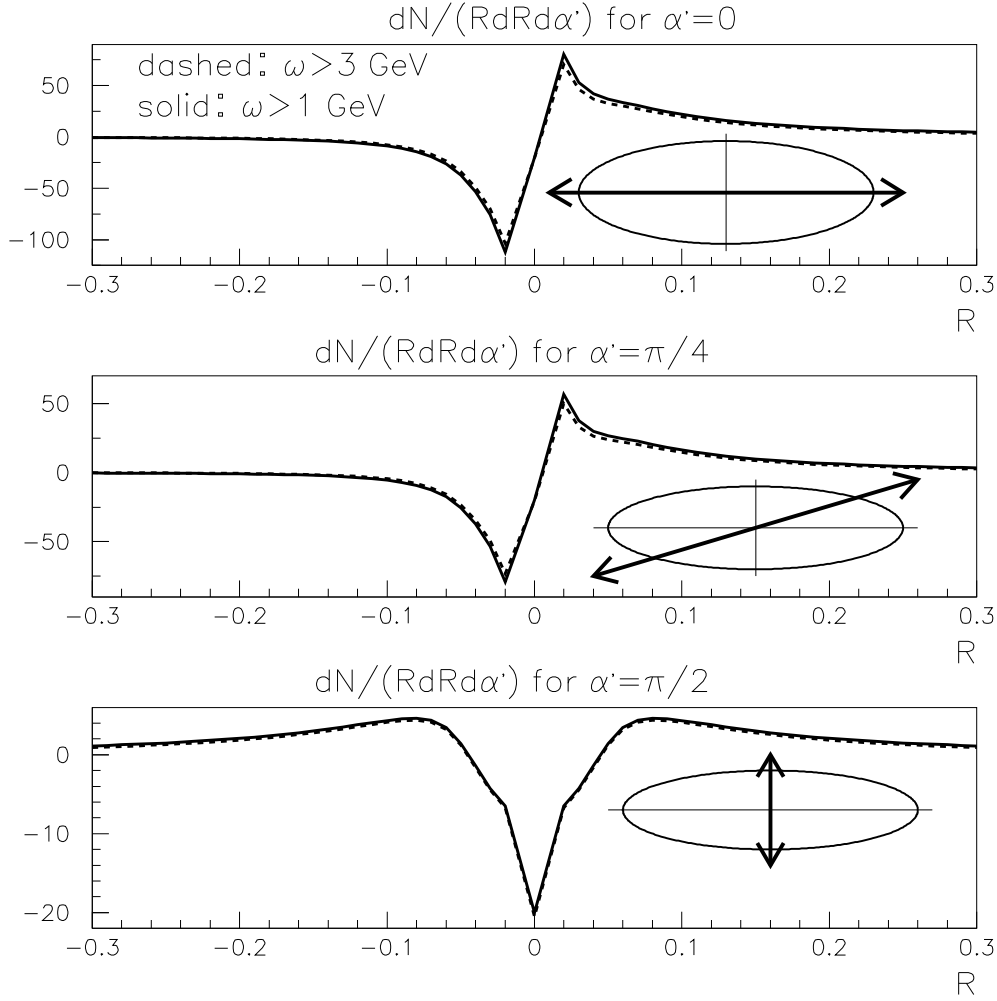}}
\vspace{0.5cm}
\caption{The medium-induced contribution (\ref{4.12}) to the total 
jet multiplicity distribution for different cuts in 
$\eta \times \phi$-plane. Parameters are the same as those of 
Fig.\protect\ref{fig6}.
}\label{fig9}
\end{figure}

If the jet energy distribution is sensitive to collective flow,
then high-$p_T$ particle correlations and jet multiplicities
should be sensitive too. Based on the medium-induced gluon energy 
distribution (\ref{2.7}), we can explore this sensitivity 
qualitatively by calculating the number of medium-induced 
gluons emitted with a gluon energy $\omega > \omega_{\rm cut}$. 
In the $\eta \times \phi$-plane,
this medium-induced multiplicity distribution reads
\begin{equation}
  N(\phi, \eta, \omega_{\rm cut})
  \equiv \frac{dN(\omega_{\rm cut})}{d\eta\, d\phi}
  = \int_{\omega_{\rm cut}}^E d\omega\, 
  \frac{dI^{\rm med}}{d\eta\, d\phi}\, .
  \label{4.12}
\end{equation}
To relate equation (\ref{4.12}) to an experimentally accessible quantity
requires a hadronization model and faces uncertainties which have
been mentioned before~\cite{Salgado:2003gb}. In Fig.~\ref{fig9}, we
plot the medium-induced modification (\ref{4.12}) of the jet 
multiplicity for the case of a collective flow in the positive 
beam direction. The qualitatively expected effects are illustrated
clearly. In the beam direction, there is a marked reshuffling 
from negative to positive rapidities due to collective flow effects. 
In the direction orthogonal to the beam, there is a somewhat smaller 
reshuffling from smaller to larger cone sizes: this is the result 
of multiple scattering in an isotropic medium which leads to a characteristic
broadening of the jet multiplicity distributions.

As a result of the eikonal approximation which underlies the calculation
of the medium-induced gluon energy distribution (\ref{2.7}), the 
leading hard parton does not change its direction due to medium effects.
This corresponds to the assumption that the leading hadron of the jet is 
located at $\eta \approx \phi \approx 0$. Then, Fig.~\ref{fig9} provides
an estimate of the rotational asymmetry of hadron production associated 
to a high-$p_T$ trigger particle. Recently, two-particle correlations 
and their possible medium-modifications have been discussed in several 
approaches~\cite{Majumder:2004br,Majumder:2004wh,Qiu:2004da,Kharzeev:2004bw,Jalilian-Marian:2004da}. 
Based on the present study, we expect flow-induced asymmetries to
affect such two-particle correlation measurements in nucleus-nucleus
collisions. In particular, asymmetries are expected in multiplicity 
distributions associated to high-$p_T$ trigger particles and in leading 
two-particle correlations~\cite{Armesto:2004pt}.
%
%
%

\section{Parton energy loss in dynamical simulations}
\label{sec5}

\subsection{A proposal to determine parton energy loss from the 
energy-momentum tensor}
\label{sec5a}

In a realistic dynamical scenario of a nucleus-nucleus collision, the
produced hard parton propagates through a medium of varying spatial
and temporal energy density $\epsilon({\bf r},z,\xi)$ and varying
collective flow $u_{\mu}({\bf r},z,\xi)$. Thus, its parton energy
loss will depend on the spatial position $\vec{r}_0$ of its
production point (the production time is $\xi \sim 0$ for hard
processes), and the orientation $\vec{n}$ of its trajectory,
\begin{equation}
  \vec{r}(\xi) = \vec{r}_0 + \xi\, \vec{n}\, .
  \label{5.1}
\end{equation}
In the absence of collective flow, numerical studies of the
medium-induced gluon energy distribution have shown that the 
medium-induced gluon radiation for a medium of time-dependent 
density is equivalent to that of a static medium whose 
density has been rescaled appropriately. This rescaling
requires the determination of the linearly line-averaged 
characteristic gluon energy along the trajectory  $\vec{r}(\xi)$ 
~\cite{Salgado:2002cd,Gyulassy:2000gk,Wang:2002ri}, 
\begin{eqnarray}
  \omega_c\left[\vec{r}(\xi)\right] 
  &=& \int_0^\infty d\xi\, \xi\, c\, \epsilon^{3/4}(\vec{r}(\xi),\xi)\, .
  \label{5.2}
\end{eqnarray}
Here, we have expressed the BDMPS transport coefficient
(\ref{1.1}) in terms of the local energy density,
$\hat{q}(\vec{r},\xi) = c\, \epsilon^{3/4}(\vec{r},\xi)$.
In the same way, one can determine the time-averaged total
transverse momentum squared
\begin{eqnarray}
   \left(\hat{q} L\right)\left[\vec{r}(\xi)\right] 
   &=& \int_0^\infty d\xi\, c\, \epsilon^{3/4}(\vec{r}(\xi),\xi)\, ,
  \label{5.3}
\end{eqnarray}
and construct the quotient 
$\omega_c\, L = 2 \frac{\omega_c^2}{\hat{q} L}$. The probability
distribution that an additional fraction $\Delta E$ of the parton 
energy is lost due to medium-induced scattering depends on 
$\omega_c$ and $\omega_c\, L$. A numerical routine for its calculation is
publicly available~\cite{Salgado:2003gb}. The characteristic 
gluon energy (\ref{5.2}) and momentum broadening (\ref{5.3}) can 
be related to the model
parameters entering the medium-induced gluon energy distribution 
(\ref{2.3}) via
\begin{eqnarray}
 \omega_c &=& \frac{1}{2} \hat{q} L^2 = \frac{1}{2} (n_0 L)\, 
               \mu^2\, L\, ,
 \label{5.4} \\
 \hat{q} L &=& (n_0 L)\, \mu^2\, .
 \label{5.5}
\end{eqnarray}
Remarkably, the time-averages (\ref{5.2}) and (\ref{5.3}) do not
require an a priori knowledge of the in-medium pathlength $L$.
In the case of a time-independent energy density of a medium
of finite size, $\epsilon({\bf r}) \propto \Theta(|{\bf r}| - L)$, 
one recovers the expressions for the static case.
For further details of how to relate parton energy loss in a
time-dependent medium to an equivalent time-independent 
calculation, we refer to 
Ref.~\cite{Salgado:2002cd,Salgado:2003gb,Baier:1998yf}.

Collective flow is an additional source of momentum
transfer to the hard parton, and will result in additional parton energy
loss. To account for this effect, we suggest to replace the energy
density in (\ref{5.2}) by the relevant boosted component of the
energy-momentum tensor (\ref{1.2}). To be more specific,
we consider the component $T^{n_\perp n_\perp}$ where $n_\perp$
is orthogonal to the trajectory (\ref{5.1}) of the hard parton,
\begin{equation}
  T^{n_\perp n_\perp} = p(\epsilon) +
  \left[ \epsilon + p(\epsilon) \right] 
  \frac{\vec{\beta}_\perp^2}{1 - \beta^2}\, .
  \label{5.6}
\end{equation}
Here, $\beta_\perp$ is the spatial component of the collective
flow field which is orthogonal to the parton trajectory. In general, 
all quantities entering (\ref{5.6}) will depend on space and time.
In the absence of flow effects, $\beta_\perp = 0$, the component
$T^{n_\perp n_\perp} = p$ determines the pressure and hence it
determines via the equation of state the energy density 
$\epsilon(p)$ entering (\ref{5.2}) and (\ref{5.3}). For finite
flow $\beta_\perp$, our proposal is to use $\epsilon(T^{n_\perp n_\perp})$
instead of $\epsilon(p)$ in evaluating the characteristic gluon
energy and momentum broadening,
\begin{eqnarray}
  \hat{q} = c\, \epsilon^{3/4}(p)\, \qquad
  \longrightarrow
  \qquad
  \hat{q} = c\, \epsilon^{3/4}(T^{n_\perp n_\perp})\, .
  \label{5.7}
\end{eqnarray}
This is consistent with what is known from analytical estimates
and numerical studies about the dependence of parton energy loss 
on momentum transfer from the medium.
For the determination of jet asymmetries in 
a dynamical scenario, relation (\ref{5.6}), one has to 
determine the relative strength of the random and directed
momentum transfers in (\ref{2.3}). For a viable model, $q_0/\mu$
should increase monotonically with 
$\frac{\epsilon + p(\epsilon)}{p(\epsilon)} 
\frac{\vec{\beta}_\perp^2}{1 - \beta^2}$. 

\subsection{Low-$p_T$ elliptic flow induces high-$p_T$ azimuthal asymmetry}

In general, a hard parton will suffer less energy loss if it 
propagates on a trajectory parallel to the flow field. Thus, for 
the same medium-induced suppression, the azimuthal asymmetry at high 
transverse momentum becomes larger when the contribution 
of the collective flow field is increased. To estimate 
the size of this effect, we consider a simple two-dimensional model. 
The hard parton is produced at an arbitrary position $(x_0, y_0)$ in the 
transverse plane according to the nuclear overlap. It propagates 
in its longitudinally comoving rest frame in the transverse direction 
$\vec{n} = \left( \cos\varphi, \sin\varphi\right)$,
along the trajectory 
\begin{equation}
  {\bf r_0}(\xi) = \left(x_0 + \xi \cos\varphi, y_0 + \xi \sin\varphi  
                    \right)\, .
  \label{5.8}
\end{equation}
For simplicity, we assume that the longitudinally comoving rest
frame of this hard parton is the longitudinal rest frame of the
medium. Then, there is only a transverse but not a longitudinal 
flow component. For the BDMPS transport coefficient which includes
collective flow effects, we make the ansatz
\begin{eqnarray}
  \hat{q}(\xi) = q_{nf} +  
                 q_{f}
                 \vert  u_T({\bf r}_0(\xi))\cdot {\bf n}_T \vert^2\, .   
  \label{5.9}
\end{eqnarray}
Here $q_f$ and $q_{nf}$ stand
for the flow and non-flow components to $\hat{q}$, the two-dimensional
vector ${\bf n}_T$ is orthogonal to the trajectory (\ref{5.8})
and projects out the corresponding transverse component
of the collective flow field $u_T({\bf r}_0(\xi))$. 
We discuss now the motivation for this ansatz. In the absence of
collective flow, $q_{nf}$ defines the 
time-averaged BDMPS transport coefficient of the dynamically
equivalent static scenario, as specified in the discussion
of (\ref{5.2}) and (\ref{5.3}). Thus, the ansatz (\ref{5.9}) 
can account for one of the main effects of longitudinal expansion, namely
the time-dependent decrease of scattering centers. In the presence of
collective flow, there is an additional momentum transfer orthogonal
to the parton trajectory and hence parallel to 
$\vec{n}_T  = \left( - \sin\varphi, \cos\varphi\right)$. Since
the BDMPS transport coefficient denotes the squared average 
momentum transfer per unit pathlength, this contribution 
enters quadratically, $\vert u_T({\bf r}_0(\xi))\cdot {\bf n}_T \vert^2$. 
For a small collective flow field,
this quadratic dependence is consistent with the more general
ansatz (\ref{5.6}). 

For an exploratory model study, we use a
blast-wave parameterization of the hadronic freeze-out stage 
of the collision~\cite{Retiere:2003kf}. 
The transverse density distribution of the produced matter is 
specified by
\begin{equation}
  \Omega(r,\phi_s) 
  = \frac{1}{1 + \exp\left( \frac{\hat{r} - 1}{a_s}\right)}\, ,
  \label{5.10}
\end{equation}
where $\hat{r} = \hat{r}(r,\phi_s)$ denotes a rescaled elliptic
position vector,
\begin{equation}
  \hat{r}(r,\phi_s) = \sqrt{ \frac{x^2}{R_x^2} + \frac{y^2}{R_y^2} }\, .
  \label{5.11}
\end{equation}
Here, $R_x$ and $R_y$ are the extension of the collision region
in the reaction plane and orthogonal to it, $\phi_s$ is the 
azimuthal angle with respect to the reaction plane, and we write
transverse positions $(x,y)$ in 
radial coordinates $x = r \cos \phi_s$, $y = r \sin\phi_s$. 
We choose a sharp, almost box-like 
density distribution with $a_s = 0.002$.
For the flow field $u_\mu(x)$, we assume longitudinal Bjorken
expansion and we use the standard notation
\begin{eqnarray}
 u_\mu(x) = \left( \cosh \eta\, \cosh \rho, \sinh \rho\, \cos \phi_b,
                   \sinh \rho\, \sin \phi_b, \sinh \eta\, \cosh \rho
 \right)\, .
 \label{5.12}
\end{eqnarray}
The coordinate $\eta$ denotes the longitudinal space-time rapidity,
and we work at mid-rapidity $\eta = 0$; 
$\phi_b$ defines the orientation which is orthogonal to the
elliptic freeze-out surface assumed in the model, 
$\tan \phi_s = \tan \phi_b \left(R_y/R_x\right)^2$. The transverse 
flow is parameterized as 
\begin{equation}
  \rho(r,\phi_s) = \hat{r} \left[ \rho_0 + \rho_a \cos(2 \phi_b)\right]\, ,
  \label{5.13}
\end{equation}
where $\rho_0 = 0.88$, $\rho_a = 0.048$ for semi-peripheral Au+Au
collisions~\cite{Retiere:2003kf}.  For the 
transverse radius parameters, we do not use the extension at 
freeze-out, but the initial transverse radii for an impact
parameter $b=7$ fm, namely $R_x = 3.1$ fm and $R_y = 5.6$ fm. 
%
\begin{figure}[h]\epsfxsize=10.7cm
\centerline{\epsfbox{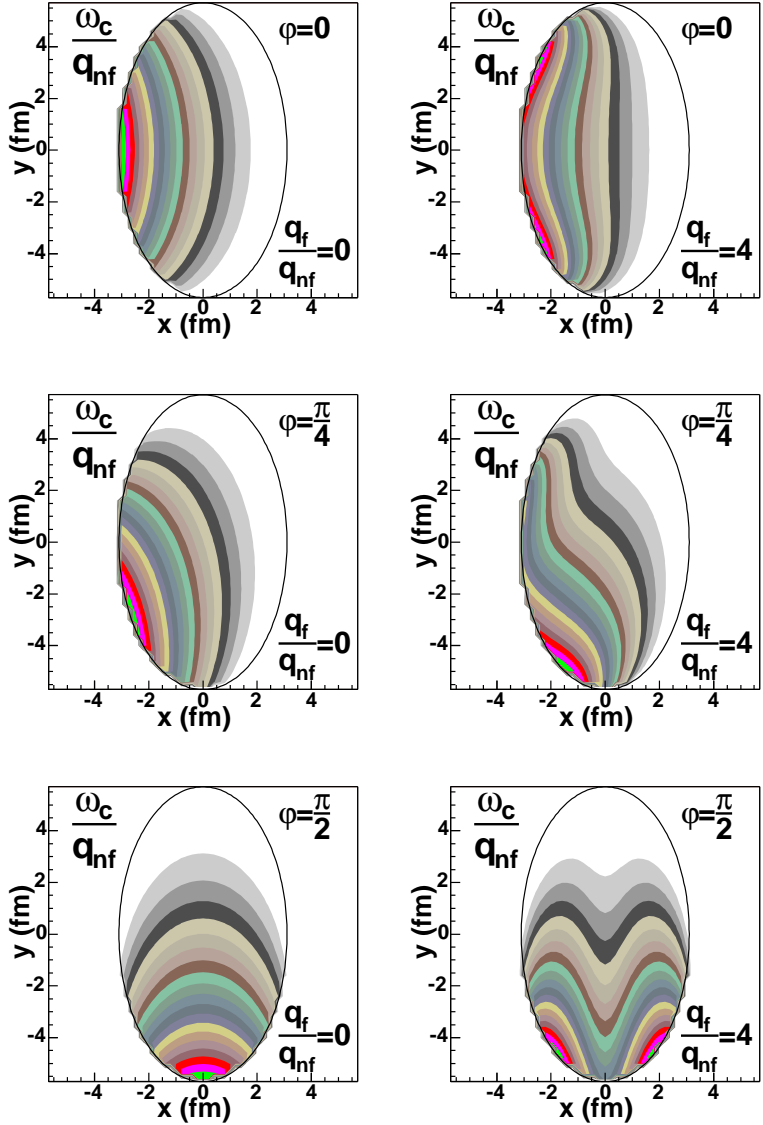}}
\vspace{0.5cm}
\caption{Contour plots of the characteristic gluon energy 
(\protect\ref{5.14}) as a function of the production point
${\bf r}_0=(x,y)$ of the hard parton and for different angles
$\varphi = 0$, $\pi/4$, $\pi/2$ of its trajectory. The
dependence of $\omega_c$ on the relative flow strength
$q_f/q_{nf}$ indicates the extent to which hard 
partons can escape with less energy loss on trajectories
parallel to the flow field. See text for more details.
}\label{fig11}
\end{figure}
%

With this input, we calculate the characteristic gluon energy
and average transverse momentum squared for a parton trajectory
(\ref{5.8}) in a medium characterized by its density distribution
(\ref{5.10}) and its collective flow field (\ref{5.12}). With the
ansatz (\ref{5.9}) for the BDMPS transport coefficient, we find
\begin{eqnarray}
  \omega_c({\bf r}_0,\varphi) &=& \int_0^\infty d\xi\, \xi\, \hat{q}(\xi)\, 
  \Omega({\bf r}(\xi), \xi)\, ,
  \label{5.14} \\
  \left(\hat{q} L\right)({\bf r}_0,\varphi) 
   &=& \int_0^\infty d\xi\, \hat{q}(\xi)\, 
  \Omega({\bf r}(\xi), \xi)\, .
  \label{5.15}
\end{eqnarray}
For a qualitative estimate of the size of parton energy loss,
one can use the pocket formula $\Delta E \approx \alpha_s 
\omega_c$~\cite{Baier:2002tc}. This motivates to investigate
$\omega_c({\bf r}_0,\varphi)$ as a function of the production
point ${\bf r}_0$ of the hard parton for different orientations
$\varphi$ of the parton trajectory. As seen from (\ref{5.9}),
$\omega_c$ depends linearly on $q_{nf}$ and
on the relative flow strength $q_f/q_{nf}$. As 
this flow strength is increased, $\omega_c$ increases for
parton trajectories which are not parallel to the flow field. 
Thus, the distortions seen in Fig.~\ref{fig11} provide a first
indication of the extent to which parton energy loss depends
on a transverse flow field and affects the azimuthal distribution
of inclusive hadron spectra. 
\vspace{-2.0cm}
\begin{figure}[h]\epsfxsize=10.7cm
\centerline{\epsfbox{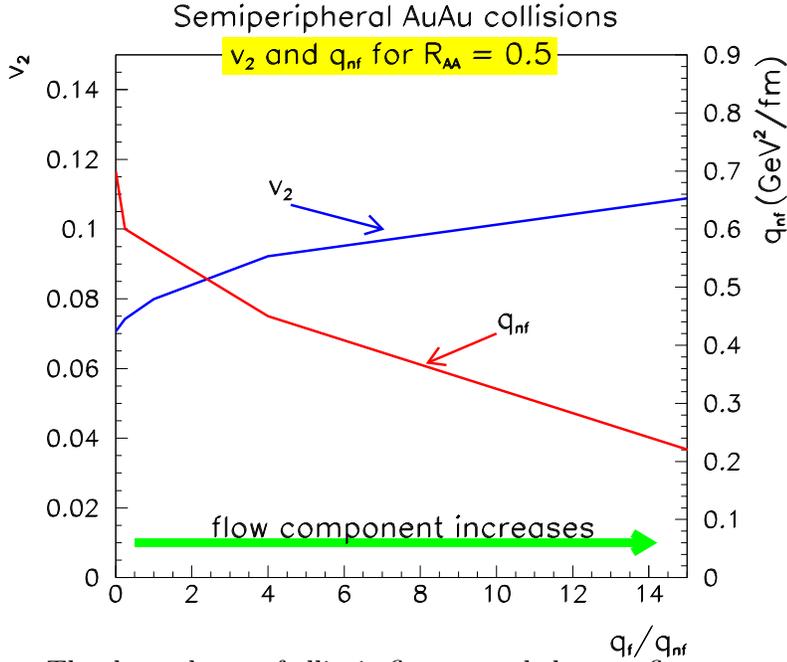}}
\vspace{-0.0cm}
\caption{The dependence of elliptic flow $v_2$ and the non-flow
component of the BDMPS transport coefficient ${q}_{nf}$
on the relative flow strength ${q}_f/{q}_{nf}$, for the
case of a nuclear modification factor $R_{AA} = 0.5$ in 
semi-peripheral Au+Au collisions. The calculation is done at
fixed transverse momentum $p_T = 7$ GeV. 
}\label{fig12}
\end{figure}
%

To estimate the effects of transverse flow, we have calculated
from (\ref{5.14}) and (\ref{5.15}) the relative suppression
of hadronic spectra due to medium-induced parton energy loss
\begin{equation}
  N(x_0,y_0,\varphi,p_T) =
  \frac{d\sigma^{med}}{dp_T} \Bigg/  
  \frac{d\sigma^{vac}}{dp_T}\, .
  \label{5.16}   
\end{equation}
Our evaluation of (\ref{5.16}) follows Ref.~\cite{Salgado:2003gb}: 
we assume a
power-law $\frac{d\sigma^{vac}}{dp_T} \propto \frac{1}{p_T^7}$,
and we calculate the medium-modification via the quenching
weights~\cite{Salgado:2003gb,Baier:2001yt,Arleo:2002kh} 
which depend on (\ref{5.14}) and (\ref{5.15}). 
The integral of (\ref{5.16}) over ${\bf r_0}$ and $\varphi$ 
weighted with the density of production points determines
the nuclear modification factor $R_{AA}$. We adjust
the non-flow component ${q}_{nf}$ such that $R_{AA} = 0.5$
which is the experimentally observed value for semi-peripheral
collisions of impact parameter $b=7$ fm. The results shown
in Fig.~\ref{fig12} were obtained for $p_T = 7$ GeV and
$\alpha_s = 1/3$. They
illustrate two qualitative effects of transverse flow: First,
low-$p_T$ elliptic flow induces an additional contribution to
high-$p_T$ azimuthal asymmetry. This effect may reduce significantly
the discrepancy of models of parton energy 
loss~\cite{Drees:2003zh,Dainese:2004te} 
in accounting for high-$p_T$ $v_2$. Second, the presence of 
collective flow diminishes strongly the local energy density
$\epsilon \propto {q}^{4/3}_{nf}$ of the medium required
for a nuclear modification factor $R_{AA}$ of fixed size.

%
\section{Conclusion}
\label{sec6}

In general, hard initially produced partons are not staying in the
locally comoving rest frame of the QCD matter generated in
a nucleus-nucleus collision. Rather, they propagate
through a matter which has collective velocity components orthogonal
to the parton trajectory. The resulting flow-induced directed 
momentum transfer can modify parton splitting significantly. 
Here, we have studied this effect by calculating the 
triple-differential medium-induced gluon energy distribution 
(\ref{2.9}) radiated off a hard parton as a function of gluon energy, 
gluon transverse momentum and azimuthal angle with respect to the 
flow field. Directed momentum transfers lead to a marked
asymmetry of the medium-induced energy distribution, since
partonic fragmentation moves significantly with the direction
of the collective flow field, see section~\ref{sec3}.

Based on the medium-induced gluon radiation spectrum (\ref{2.3}),
and simple assumptions about the dynamical evolution of the
matter produced in nucleus-nucleus collisions, we have reached
several qualitative conclusions of phenomenological relevance. 
In particular, as discussed in section~\ref{sec4}, flow-induced 
distortions of parton fragmentation will be experimentally
accessible in calorimetric jet measurements, multiplicity 
distributions associated to high-$p_T$ trigger particles,
and leading two-hadron correlation functions. Moreover, 
as seen in Fig.~\ref{fig12}, different combinations of local 
energy density and collective flow can account for the same
suppression of single inclusive hadron spectra. This illustrates
the generic argument of section~\ref{sec5a} that the strength 
of parton energy loss is not governed by the local energy density, 
but rather by the local energy-momentum tensor (\ref{1.2}).
Flow effects can also contribute appreciably to the size of 
high-$p_T$ $v_2$ which is underpredicted in recent
model comparisons~\cite{Drees:2003zh,Dainese:2004te}.

In general, the effects of medium-induced parton energy loss 
depend on time-integrated properties of the medium, see 
section~\ref{sec5a}. Thus, a more quantitative study 
of flow-induced parton energy loss effects requires a realistic
model of the dynamical evolution of the collision region. It
also requires information about the spatial distribution of hard
processes in the produced matter. Determining this information is
a challenge which 
- in an interplay of theory and further data analysis at RHIC
and the LHC - should come within reach in the near future. 
We hope that our work is of use for further studies in this
direction and in particular for relating the dynamics of a 
hydrodynamical medium to the dynamics of hard processes in that
medium, started in Ref.~\cite{Hirano:2002sc,Hirano:2003pw}. 

{\bf Acknowledgment:} We thank Rudolf Baier, Peter Jacobs, Dan Magestro,
Andreas Morsch, J\"urgen Schukraft and Fuqiang Wang for helpful discussions.


\end{document}